\documentclass{article}

\usepackage[final, nonatbib]{neurips_2025}

\usepackage[utf8]{inputenc} %
\usepackage[T1]{fontenc}    %
\usepackage{booktabs}       %
\usepackage{amsfonts}       %
\usepackage{nicefrac}       %
\usepackage{microtype}      %
\usepackage{xcolor}         %
\usepackage{colortbl}
\usepackage{multirow}
\definecolor{fade}{gray}{0.4}
\definecolor{green}{rgb}{0.0, 0.5, 0.0}
\definecolor{grey}{rgb}{0.5, 0.5, 0.5}

\usepackage{soul}

\usepackage{fontawesome}

\usepackage{pifont}%
\usepackage{graphicx}
\usepackage{amsmath}
\usepackage{booktabs}
\usepackage{enumitem}
\usepackage{color, colortbl}
\usepackage{multirow}
\usepackage{diagbox}
\usepackage{algorithm}
\usepackage{listings}
\definecolor{rc1}{RGB}{235,235,235}
\definecolor{rc2}{RGB}{255,255,255}
\definecolor{codeblue}{rgb}{0.25,0.5,0.5}
\definecolor{codekw}{rgb}{0.85, 0.18, 0.50}
\usepackage{booktabs}       %
\usepackage{amsfonts}       %
\usepackage{nicefrac}       %
\usepackage{microtype}      %
\usepackage{graphicx}
\usepackage{tikz}
\usepackage{comment}
\usepackage{amsmath,amssymb} %
\usepackage{color}

\usepackage{stackengine}

\usepackage{algpseudocode}

\definecolor{green}{rgb}{0.0, 0.5, 0.0}

\usepackage{caption}
\usepackage{subcaption}

\usepackage[most]{tcolorbox}
\usepackage{mdframed}
\usepackage{wrapfig}

\usepackage{bbm}
\usepackage{fix-cm} 

\usepackage{xtab}

\usepackage{alltt,xcolor}
\usepackage{relsize}
\usepackage{rotating} 
\usepackage{amsmath,amssymb,amsthm}

\newcounter{qcounter}

\usepackage{verbatim}

\newtcblisting[auto counter,number within=chapter]{sourcecode}[2][]{
  sharp corners,
  breakable,
  fonttitle=\bfseries,
  colframe=black!70,
  listing only,
  listing options={
    basicstyle=\ttfamily\scriptsize,
    language=php,
    showstringspaces=false,
    breakatwhitespace=true,
    breaklines=true,
    postbreak=\mbox{},
    tabsize=2,
    linewidth=\linewidth,
    breakindent=0pt
  },
  title=#2,
  #1
}

{
    \begin{tcolorbox}[
        width=#1, 
        colframe=black, 
        colback=gray!10, 
        sharp corners, 
        boxrule=0.5mm, 
        left=5pt, right=5pt, top=5pt, bottom=5pt, 
        fontupper=\ttfamily 
    ]
    \small
}
{
    \end{tcolorbox}
}

\usepackage{amsmath,amsfonts,bm}

\def\eqref#1{equation~\ref{#1}}

\def\1{\bm{1}}

\DeclareMathAlphabet{\mathsfit}{\encodingdefault}{\sfdefault}{m}{sl}
\SetMathAlphabet{\mathsfit}{bold}{\encodingdefault}{\sfdefault}{bx}{n}

\usepackage{url}
\usepackage[pagebackref=true,breaklinks=true,colorlinks,bookmarks=false]{hyperref}
\hypersetup{
    colorlinks=true,
    filecolor=magenta,      
    urlcolor=magenta,
}

\title{
Multi-View Hierarchical Representation Learning of Fetal Hemodynamics for Maternal Hypertension Detection at the Edge
}

\author{Alireza Rafiei \\
Emory University, Atlanta, GA, USA \\
\texttt{alireza.rafiei@emory.edu}
\And
Anahí Venzor Strader \\
Center for Indigenous Health Research, Wuqu' Kawoq | Maya Health Alliance, Tecpán, Guatemala \\
\texttt{anahivenzor@wuqukawoq.org}  
\And
Esteban Castro Aragón \\
Center for Indigenous Health Research, Wuqu' Kawoq | Maya Health Alliance, Tecpán, Guatemala \\
\texttt{estebancastro@wuqukawoq.org}
\And
Victoriana Rosibely Sut Serech \\
Center for Indigenous Health Research, Wuqu' Kawoq | Maya Health Alliance, Tecpán, Guatemala \\
\texttt{rosibely@wuqukawoq.org}
  \And
Enma Carolina Coyote Ixen \\
Center for Indigenous Health Research, Wuqu' Kawoq | Maya Health Alliance, Tecpán, Guatemala \\
\texttt{enmacoyote@wuqukawoq.org}
\And
Reza Sameni \\
Emory University \& Georgia Institute of Technology, Atlanta, GA, USA \\
\texttt{rsameni@dbmi.emory.edu}
\And
Peter Rohloff \\
Center for Indigenous Health Research, Wuqu' Kawoq | Maya Health Alliance, Tecpán, Guatemala \\
\texttt{peter@wuqukawoq.org}
\And
Gari D. Clifford \\
Emory University \& Georgia Institute of Technology, Atlanta, GA, USA \\
\texttt{gari@gatech.edu}
\And
Nasim Katebi \\
Emory University, Atlanta, GA, USA  \\
\texttt{nkatebi@emory.edu}}

\begin{document}

\maketitle

\setcounter{footnote}{0}

\begin{abstract}
Hypertensive disorders of pregnancy remain a leading cause of maternal and fetal morbidity worldwide, yet diagnosis relies on intermittent cuff-based blood pressure measurements that are prone to bias and fail to capture continuous physiological dynamics. Growing evidence suggests that fetal cardiovascular activity is associated with maternal-placental hemodynamics and may encode markers of maternal hypertension. To analyze this, we collected a large-scale dataset of fetal one-dimensional Doppler ultrasound recordings paired with maternal blood pressure from 3,255 pregnant women across 8,170 antenatal visits in rural Guatemala. We developed AutoHyPE, a hierarchical attention network that models short- and long-term signal structure, incorporating a novel prototype-based contrastive learning and multi-view strategy to enhance representation robustness under long-tailed class distribution and biological variability. AutoHyPE achieved an AUROC of 0.80 for maternal hypertension detection, outperforming baseline approaches while maintaining balanced performance across classes, with no performance degradation in an edge deployment scenario. Our findings demonstrated that fetal cardiac mechanical activity contains hemodynamic features indicative of maternal hypertension status. This supports a promising paradigm shift toward continuous, objective monitoring of maternal health using existing, low-cost ultrasound technology and introduces a complementary approach to traditional methods based on blood pressure measurements, advancing scalable prenatal care.
\end{abstract}

\section{Introduction}
\label{sec:intro}

Hypertensive disorders of pregnancy, encompassing gestational hypertension and preeclampsia, are the most common medical complications encountered during pregnancy and remain a major global obstetric challenge \cite{poon2023hypertensive,lailler2024early}. In the United States, these conditions complicate 13–15\% of pregnancies \cite{ford2022hypertensive,radparvar2024hypertensive}. Globally, although the age-standardized incidence has modestly declined, the absolute number of affected pregnancies increased from about 16.3 million to 18.1 million over the past three decades, accounting for an estimated 27,800 deaths in 2019 \cite{wang2021epidemiological}. Adverse outcomes related to hypertensive disorders of pregnancy pose serious risks to both mother and fetus, being associated with placental abruption, preterm delivery, fetal growth restriction (FGR), stillbirth, and maternal death due to stroke or eclampsia \cite{traub2024hypertensive,bisson2023preeclampsia}. Beyond the immediate perinatal period, they are increasingly recognized as harbingers of short- and long-term cardiovascular and cerebrovascular risk. For instance, women with a history of hypertension face approximately a 1.7-fold higher risk of stroke later in life \cite{brohan2023hypertensive}. Moreover, offspring of mothers with hypertension exhibit elevated blood pressure, poorer cognitive outcomes, and a greater susceptibility to renal dysfunction \cite{pinheiro2016hypertensive,turbeville2020preeclampsia}. These findings underscore the need for early detection and intervention, not only to prevent acute obstetric complications but also to mitigate long-term multisystem sequelae.

Abnormal placentation is a central pathophysiological event underlying hypertensive disorders of pregnancy, which manifests clinically with maternal hypertension and multiorgan dysfunction \cite{furuya2008pathophysiology,braunthal2019hypertension}. A key component of managing this is timely identification of hypertension, particularly in women at high risk of adverse outcomes such as early-onset preeclampsia, and undertaking the necessary actions to improve placentation and reduce disease prevalence \cite{poon2014early}. Hypertension is generally defined as two or more systolic blood pressure (SBP) readings $\geq$ 140 mmHg or diastolic blood pressure (DBP) readings $\geq$ 90 mmHg, though these thresholds have been debated and refined over time \cite{brown2018hypertensive,cifkova2023hypertension}. The current diagnostic standard for hypertension in pregnancy relies almost exclusively on intermittent cuff-based blood pressure measurements and the detection of proteinuria. However, these approaches have several inherent limitations. Cuff-based blood pressure measurements are susceptible to inaccuracies due to cuff size, patient posture, and movement, and manual devices are highly operator dependent, all of which can lead to measurement errors \cite{berg20194,ishigami2023effects}. They also present anatomical and population-specific limitations, as readings can be unreliable in women with obesity, edema, or increased vascular stiffness, and many devices have not been validated to perform reliably across diverse pregnant or preeclamptic populations \cite{bello2018accuracy}. Most importantly, they provide only discrete and momentary assessments, which fail to capture transient variations or dynamic fluctuations in blood pressure.

Advances in mobile health platforms and wearable technologies are transforming maternal–fetal health monitoring by enabling affordable, real-time, and continuous physiological assessment throughout pregnancy. When integrated with smartphone-based data acquisition and edge–cloud analytics, these systems allow remote and longitudinal evaluation everywhere pregnancy occurs, improving accessibility in resource-limited settings and supporting data-driven clinical decision-making. Of note, leveraging one-dimensional Doppler ultrasound (1D-DUS) signals augmented with automated algorithms and machine learning (ML) models through these platforms enables the assessment of fetal cardiac mechanical activity and provides clinically meaningful biomarkers predictive of maternal and fetal health risks \cite{valderrama2020review,ramos2024mobil}. This non-invasive approach captures dynamic blood flow information, offering a detailed view into fetal heart function, cardiac wall and valve motion, and overall cardiovascular status of the fetus. Prior works developed and validated different algorithmic models to extract key insights from fetal 1D-DUS data \cite{katebi2020unsupervised,rafiei2025auto,motie2025real,rafiei2025autofhr}, and a mobile health platform was designed to support the practical application of these models in real-world settings \cite{katebi2024edge}. The models encompassed various fetal–maternal monitoring tasks, including fetal heart rate estimation \cite{valderrama2019open}, fetal heart rate segmentation and variability analysis \cite{rafiei2025next}, and gestational age estimation \cite{katebi2023hierarchical}. Building on this foundation, the present work sought to further advance the capabilities of this versatile platform for broader use by exploring the potential of detecting hypertensive disorders during pregnancy using 1D-DUS signals.

Previous studies have found that fetal cardiac development and fetal heart rate variability (FHRV) are directly influenced by hypertensive disorders of pregnancy. Yum et al. \cite{yum2004instability} analyzed the instability and frequency-domain variability of fetal heart rate in pregnancies with preeclampsia, including subgroups with and without FGR. They observed that low- and high-frequency powers were significantly higher in fetuses affected by preeclampsia without FGR compared with the control cohort. 
Lakhno \cite{lakhno2017autonomic} studied the effect of preeclampsia on FHRV metrics, presenting that autonomic imbalance measured through modulated cardiotocographic (CTG) variables captured both maternal and fetal circulatory responses to preeclampsia and reflected the progression toward fetal distress. In a related investigation, Lakhno \cite{lakhno2014impact} examined fetal electrocardiology (FECG) morphology and FHRV, reporting that increasing disease severity was associated with shortened PQ and QT intervals and an elevated T/QRS ratio, indicative of altered cardiac conduction and autonomic regulation in utero. Lucero-Orozco et al. \cite{lucero2024analysis} characterized fetal autonomic alterations during the latent phase of labor, showing that fetuses of women with preeclampsia exhibited reduced FHRV complexity and increased low-frequency power, reinforcing the concept of impaired fetal autonomic modulation in hypertensive pregnancies. Complementary evidence has emerged from studies of fetal cardiac structure. Youssef et al. \cite{youssef2020fetal} observed that pregnancies complicated by either preeclampsia or FGR demonstrated similar patterns of fetal cardiac remodeling, including increased ventricular sphericity, greater relative wall thickness, and evidence of myocardial dysfunction reflected in increased myocardial performance index and cardiac biomarkers. Huluta et al. \cite{huluta2023fetal} studied fetuses at mid-gestation whose mothers later developed pre-eclampsia and found subtle reductions in left ventricular myocardial deformation indices, suggesting that fetal cardiac programming may begin even before overt maternal disease manifests. Similarly, Zhang et al. \cite{zhang2025effects} demonstrated that fetuses of pregnancies with hypertensive disorder exhibit subtle myocardial remodeling characterized by smaller right ventricular dimensions and reduced strain indices, with the right ventricle appearing particularly vulnerable to hypertensive exposure. Longitudinal analyses further indicate that these alterations may persist beyond birth. The study by Aye et al. \cite{aye2020prenatal} revealed that term-born infants from hypertensive pregnancies maintained smaller right ventricular end-diastolic volumes and modestly increased left ventricular mass through three months of postnatal life, findings that parallel those of Timpka et al. \cite{timpka2016hypertensive}, who reported greater left ventricular relative wall thickness and smaller end-diastolic volumes in adolescent offspring of hypertensive pregnancies. Collectively, these studies delineated a continuum in which maternal hypertensive disorders induced autonomic dysregulation and alterations in fetal cardiac structure and rhythm. In parallel, clinical studies reinforced the diagnostic utility of 1D-DUS signals for detecting preeclampsia, with discriminative parameters derived from uterine and uteroplacental waveforms \cite{franco2015using}. Multiple indices extracted from Doppler, such as the resistance index, pulsatility index, and systolic/diastolic ratio of the uterine artery, along with fetal heart rate responses and uteroplacental flow measures, were found valuable in identifying hypertensive pathology and were linked to placental perfusion \cite{franco2015using,riknagel2016digital,hoyer2017monitoring}.

Motivated by these observations, we hypothesized that the mechanical activity of the fetal heart (recorded via 1D-DUS) encoded physiologically meaningful patterns that were influenced by, and in turn reflected, maternal hypertensive states. The bidirectional coupling between maternal and fetal circulatory systems suggested the presence of trackable signatures capable of signaling elevated maternal blood pressure. To evaluate this hypothesis, we assembled a large dataset of paired 1D-DUS waveforms and maternal blood pressure measurements collected during routine prenatal visits. Leveraging this dataset, we designed a hierarchical attention network (HAN) to extract short-term cardiac signatures and longer-range, stable temporal patterns associated with hypertension. Learning in this setting posed a significant challenge due to severe class imbalance. This was further complicated by the U-shaped distribution of the input data, where most representative input values clustered at the extremes rather than a central tendency. Simply increasing model complexity in such cases could lead to overfitting, while overly constrained models failed to capture critical structure. We found that incorporating complementary views derived from transformed versions of the input substantially mitigated these issues. This encouraged the network to learn features that remained discriminative across combined views, acting as an implicit form of regularization that reduced reliance on representation-specific patterns. Additionally, a novel prototype-based contrastive formulation was used to pretrain the network to organize embeddings around semantic class centers and provided a more stable learning landscape. Detailed analysis and ablation studies demonstrated the impact of each component of our solution. In summary, our contributions are as follows:
\begin{itemize}
\item We collected, curated, and processed a dataset comprising fetal 1D-DUS recordings and blood pressure measurements from pregnant women.

\item We proposed AutoHyPE, a deep HAN approach utilizing window- and sequence-level encoders to decipher maternal hypertensive patterns within fetal 1D-DUS signals.

\item We introduced a multi-view learning framework combined with prototype-based contrastive pretraining to mitigate class imbalance and bimodal boundary input distribution while stabilizing the learning process.

\end{itemize}

\section{Materials and Methods} \label{sec:method}
\subsection{Data Collection}
We collected and assembled a multimodal dataset comprising paired fetal 1D-DUS waveforms and maternal blood pressure measurements recorded during routine antenatal encounters as part of a randomized controlled trial and a subsequent follow-up observational cohort study conducted in rural highland Guatemala near Tecpán, Chimaltenango. To our knowledge, this is the first study to prospectively acquire 1D-DUS recordings and blood pressure values at scale, enabling direct investigation of how fetal hemodynamic signatures reflect maternal hypertensive states. Data were gathered from 3,255 pregnant women during 8,170 visits spanning different stages of gestation. A network of 121 trained Indigenous midwives participated in the data collection process as part of an ongoing maternal–fetal health monitoring initiative and followed standardized acquisition protocols; more details regarding midwife training, proficiency assessment, and study procedures are available in \cite{martinez2018mhealth} and \cite{stroux2016mhealth}. 
The study was approved by the Institutional Review Boards of Emory University, Wuqu’ Kawoq | Maya Health Alliance, and Agnes Scott College (protocol numbers: IRB00076231, WK-2015-001, and 02.02.2015, respectively), registered as a clinical trial (ClinicalTrials.gov, identifier: NCT02348840), and all data were fully de-identified before analysis.

Across visits, midwives recorded the 1D-DUS signals in a supine position using a handheld AngelSounds JPD-100S device (Jumper Medical Co., Ltd., Shenzhen, China) operating at a transmission frequency of 3.3\,MHz. The device was connected to a smartphone through a custom audio cable incorporating a capacitor–resistor network that emulated the electrical behavior of a standard hands-free headset input. A small external speaker was also attached, allowing midwives to perform real-time audio checks and confirm that the probe was correctly positioned by listening for audible fetal heartbeats. Recordings were stored as 16-bit, uncompressed WAV files sampled at 44.1\,kHz. We developed a dedicated mobile application to compile essential clinical information and to guide midwives through the acquisition workflow and related tasks \cite{katebi2024edge}.

Immediately before each Doppler recording, midwives entered key clinical metadata, including gestational age estimated from the last menstrual period, into the mobile application. Maternal blood pressure was then measured on both arms using a validated automatic oscillometric monitor (Omron M7, Omron Co., Kyoto, Japan), a device previously confirmed to perform reliably in both normotensive and preeclamptic pregnancies \cite{bello2018accuracy}. All images were subsequently processed using a semi-automated transcription pipeline designed to identify the device’s LCD display region and extract the SBP and DBP values \cite{kulkarni2021cnn, katebi2024automated}. The extracted measurements were then paired with the corresponding 1D-DUS waveform and associated metadata, forming a complete multimodal record for each visit to enable consistent acquisition of high-quality physiological data, even in remote settings with limited resources. Table \ref{tab:ga_bb} summarizes the mean and standard deviation of blood pressure measurements across crude, month-level gestational age estimates, and Figure \ref{fig:iad} illustrates the inter-arm differences observed in the bilateral blood pressure recordings obtained during the visits.

\subsection{Data Preparation}
In alignment with previous research on fetal Doppler signal analysis and cardiotocography, as well as with automated algorithms developed for the fetal–maternal mobile health monitoring platform, all 1D-DUS recordings were segmented into 3.75-second windows and resampled to 4\,kHz \cite{rafiei2025auto,motie2025real,katebi2024edge}. This standard configuration provides an effective balance between data sufficiency and signal stationarity for physiological analysis \cite{valderrama2019open}. Each 3.75-second window was formed by combining five consecutive 0.75-second segments of the signals. Of note, 
to enhance data utilization and preserve boundary information, a 0.75-second stride was applied between consecutive segments, allowing partial overlap among neighboring windows. This overlapping design ensured that transient yet diagnostically relevant signal dynamics near segment edges were not lost.

Because 1D-DUS signals are inherently susceptible to noise and artifacts, their quality was assessed prior to model development and validation. Potential sources of quality degradation included fetal or maternal motion, ambient environmental noise, and inconsistent transducer placement. To address this, we utilized a high-performance signal quality assessment model to identify and exclude low-quality windows from each recording, as described and detailed in \cite{motie2025real}.

\begin{figure}[ht]
    \centering
    \includegraphics[width=0.5\linewidth]{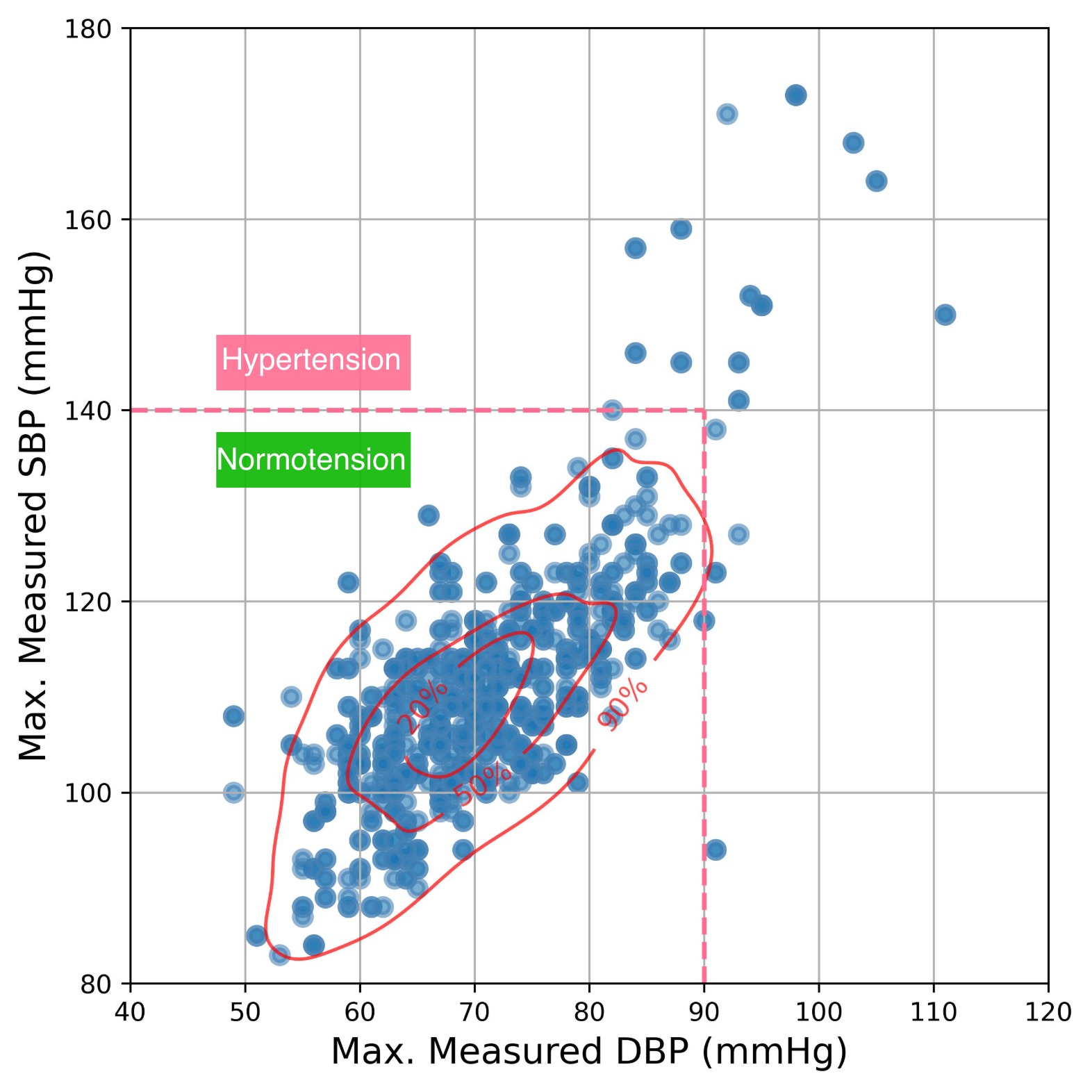}\\
    \caption{Joint distribution of the maximum SBP and DBP measurements obtained across both maternal arms for all paired 1D-DUS signals and blood-pressure recordings. Multiple points may overlap, resulting in stacked markers at the same coordinate. Dashed lines denote the clinical thresholds for hypertension (SBP $\geq$ 140 mmHg or DBP $\geq$ 90 mmHg). Samples above or to the right of these thresholds are classified as hypertensive, whereas those below both thresholds are classified as normotensive. Contours represent kernel density estimates, with labeled levels indicating the percentage of observations contained within each contour.}
    \label{fig:sbp}   
\end{figure}

Following segmentation and quality screening, each 3.75-second window was transformed into a time–frequency representation using the continuous wavelet transform (CWT) with a Morlet mother wavelet. The resulting scalogram captured the temporal evolution of frequency components across the fetal cardiac cycle, providing a two-dimensional representation with superior localization of transient features. From these wavelet coefficients, a reciprocal scalogram was also computed. Both the original scalogram and its reciprocal reconstruction were then normalized separately to their respective minimum and maximum amplitude values to mitigate inter-subject variability and device-dependent amplitude differences. The analysis in this study focused on recordings obtained during the ninth month of pregnancy (827 total visits), as hypertensive disorders most commonly manifest in late gestation, when vascular resistance and hemodynamic alterations peak. 
For visits with available blood pressure measurements, the final processed dataset comprised 742 normotensive and 33 hypertensive paired 1D-DUS recordings. Maternal hypertension status was determined using the maximum recorded SBP and DBP values measured across both arms, a method recommended in clinical guidelines to avoid underdiagnosis, since inter-arm blood-pressure differences are common and may reflect clinically important vascular variation \cite{clark2022higher, poon2008inter}. The joint distribution of maximum SBP and DBP values for all cases in this cohort is shown in Figure \ref{fig:sbp}.

\subsection{Proposed Approach}

\paragraph{Multi-view Learning Strategy.}
For deep learning tasks involving waveform inputs, such as audio recordings or high-frequency physiological signals, the analysis pipeline typically begins by converting raw signals into time–frequency representations (e.g., a scalogram or spectrogram). These representations inherit the statistical peculiarities of the underlying signals; however, they sometimes display long-tailed intensity distributions, pronounced energy asymmetries, and large regions of near-zero power. Such characteristics can limit the expressiveness of neural network architectures and bias attention mechanisms toward high-amplitude regions, especially in label-imbalanced settings where model depth should be constrained to avoid overfitting.

To address this, we introduced a reciprocal feature augmentation (RFA) strategy that provided the network with a multi-view representation comprising both the standard scalogram $S(t,f)$ and its deterministic reciprocal transformation, defined as $g(S(t,f)) = \frac{1}{S(t,f) + \epsilon}$. While this transformation is invertible and therefore information-preserving, it profoundly reshaped the geometry of the input space. Specifically, it redistributed variance across the domain, altered local Lipschitz constants, and modified the pairwise distances between inputs. Since neural network components, such as filters, activation functions, and poolings, are not invariant to such monotonic intensity transforms, the mapping $g(S(t,f))$ offers an alternative embedding of the same underlying manifold, one whose structure may be more congenial for the network to model and to the inductive biases of the architecture. A key insight here was that the high-energy structures dominated the scalogram, reflecting clutter, strong flow bands, and abrupt spectral shifts, naturally exerted a disproportionate influence on early convolutional filters. Meanwhile, other potential diagnostically important cues, such as spectral windows, signs of turbulence, and weak harmonic content, resided in low-energy regions. The reciprocal scalogram counterbalanced this bias. Because its gradient was largest for small values of $S(t,f)$, it amplified faint structures in regions where the scalogram provided little learning signal. Subtle spectral troughs and weak flow signatures, which previously contributed negligible gradients, became salient and discoverable. The proposed two-channel representation $[S(t,f), g(S(t,f))]$ thus enabled the network to leverage complementary coordinate systems over the same signal to improve feature separability and stabilize gradient propagation across the full dynamic range of the time–frequency representation. Because the two channels were deterministically related yet differently weighted, the model was implicitly encouraged to learn view-consistent features that acted as an effective form of regularization, stabilizing the optimization dynamics and expanding the range of useful gradients available to early layers. Comparative examinations of characteristics of the scalogram and its reciprocal view across hypertensive and normotensive groups revealed distinct structural behaviors that were not apparent when the signal was viewed through a single representation (Appendix. \ref{sec:char}). Analyzing these views showed clear contrasts in how spectral complexity, variability, and distributional patterns evolved across the two groups (Figures \ref{fig:3d_kdp} and \ref{fig:entropy}).

\paragraph{Network Implementation.}
To effectively capture both the short- and long-term patterns of fetal cardiac activity, we designed a HAN inspired by the gestational age estimation model introduced in \cite{katebi2023hierarchical}. The architecture was developed to jointly learn morphological and temporal representations of 1D-DUS signals, enabling the network to focus on diagnostically relevant dynamics associated with maternal hypertensive status. The HAN model operated at two levels of abstraction: the window level, which processed each 3.75-second segment, and the sequence level, which modeled temporal relationships across consecutive windows within a recording. This dual-level structure allowed the network to capture both (1) local dynamic patterns of successive fetal heartbeats and (2) broader contextual variations that evolved over time. Such variations may arise from changes in fetal activity, differences in noise characteristics and motion-related artifacts, or shifts in the relative importance of specific cardiac phases.

\begin{figure}[ht]
    \centering
    \includegraphics[width=\linewidth]{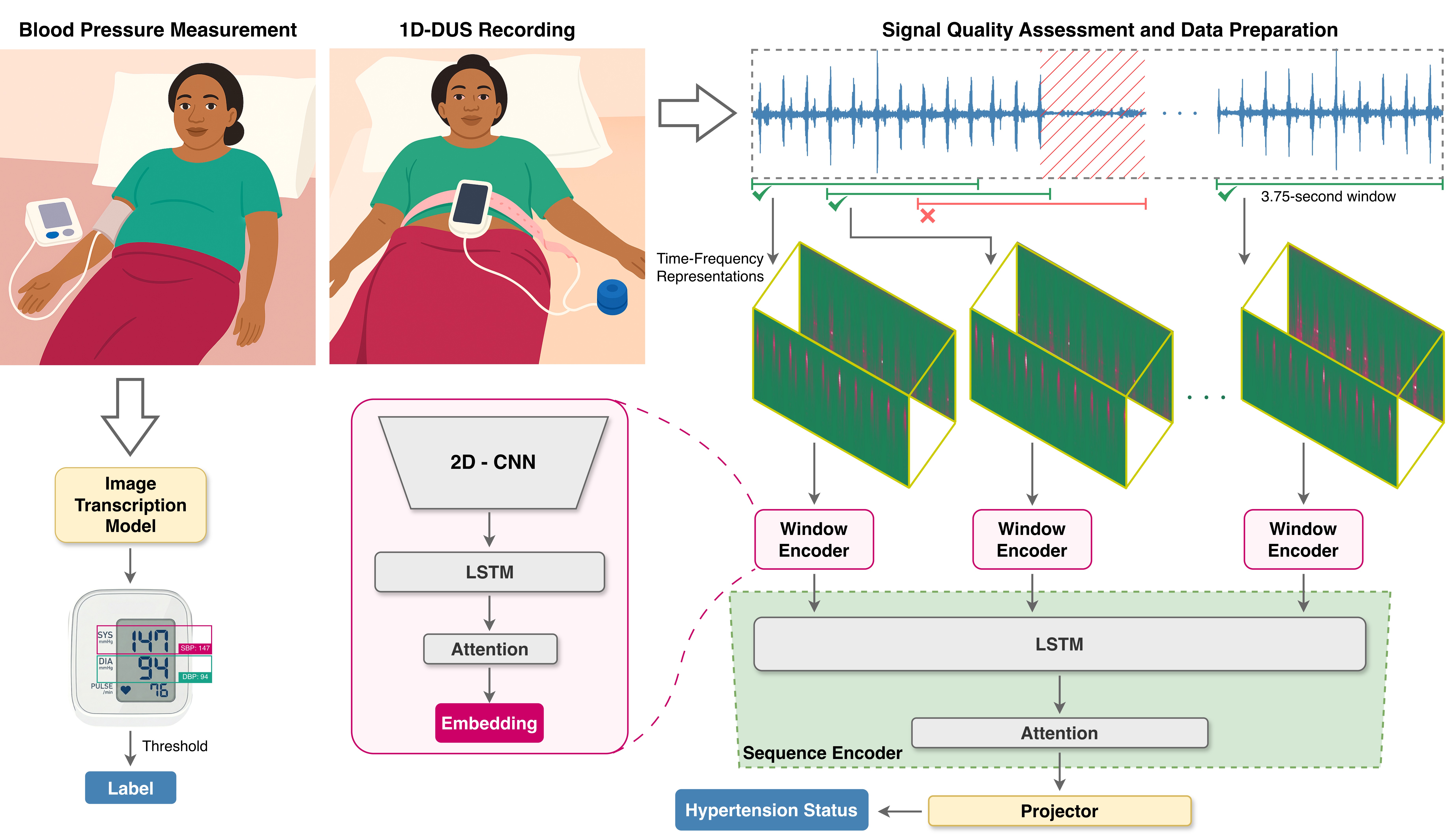}\\
    \caption{Overview of the proposed framework for automated maternal hypertension detection using fetal 1D-DUS signals. The pipeline begins with data acquisition: blood pressure measurements are digitized via an image transcription model to generate ground truth labels, while 1D-DUS recordings undergo signal quality assessment. Signals are then segmented into 3.75-second windows and converted into multi-view time-frequency representations. These inputs are processed by a HAN architecture comprising a window encoder and a sequence encoder, which aggregates temporal features to assess hypertension status.}
    \label{fig:main}   
\end{figure}

At the window level, we employed an encoder module to extract features from each of the ten individual multi-view windows of the input data. This module utilized a convolutional backbone consisting of three sequential blocks, each comprising a convolutional layer, batch normalization, spatial dropout, and max-pooling. The resulting feature maps of these blocks were then passed to an LSTM layer that aggregated the within-window structure into a higher-level temporal descriptor, after which a window-level soft attention layer produced a contextualized representation for each 3.75-second window. At the sequence level, another encoder integrated the series of embeddings from all windows and used an LSTM layer to learn longer-range, inter-window temporal patterns. This was followed by an attention layer that explicitly modeled relationships across windows, assigning adaptive weights based on their relative contribution, thereby emphasizing informative patterns. Finally, a projector refined these high-level features to ensure they were optimally formatted for the final classification layer. Figure \ref{fig:main} depicts a schematic overview of the proposed framework’s workflow.

\paragraph{Learning Objective.}
We assessed both the direct supervised performance of the developed HAN model and the benefits of enhancing its encoders through contrastive representation learning. As a baseline, the full HAN architecture was trained in a standard end-to-end supervised fashion to predict the likelihood of maternal hypertension from the input recording. Given the substantial variability in 1D-DUS signals, we reasoned that richer and more stable representations could be learned by pretraining the encoders using contrastive learning-based objectives prior to the classification task. This approach enforces a mapping that pulls normalized embeddings of the same class closer together and separates those belonging to different classes.

To investigate this, we explored two main contrastive strategies: a pairwise contrastive loss (CL) and a prototype-based contrastive loss (PCL). The CL objective followed a classical supervised contrastive framework, where embeddings were normalized, compared through a temperature-scaled cosine similarity matrix, and optimized with an N-pairs loss to pull together samples sharing the same label. In contrast, the proposed PCL strategy adopted a parametric, prototype-driven formulation, where the model learned trainable class prototypes residing in the same embedding space as latent 1D-DUS features. Unlike pair-wise approaches, the extracted features were contrasted directly against these prototypes via a temperature-scaled softmax over cosine similarities. This PCL formulation yielded an objective that encouraged embeddings to cluster around their corresponding class prototype while separating from the prototype of the opposite class. Unlike prior approaches, the prototypes in PCL were free parameters jointly learned with the encoder, making this formulation distinct. Practically, the PCL approach resulted in several advantages, including reduced sensitivity to batch composition, improved robustness in the presence of noisy or heterogeneous data, lower computational cost, and more stable class-level embedding structures.

\subsection{Other Developed Methods and Baselines}  \label{sec:other_models}
Since no prior work was available for direct comparison with our approach, to our knowledge, we developed a set of architectural and representation learning approaches for comprehensive benchmarking. From an architectural perspective, we explored multiple configurations, including the HAN model without recurrent modules, variants that omit window-level and hierarchical attention mechanisms, and models that exclude window-level or sequence-level encoders. We further evaluated a version without convolutional feature extractors, as well as a collapsed hierarchy variant in which all windows were merged into a single long sequence. Established state-of-the-art time-series models, such as InceptionTime \cite{IsmailFawaz2020inceptionTime} and ROCKET \cite{dempster_etal_2020}, were implemented and evaluated.

In addition, we assessed several contrastive and metric-learning objectives beyond the CL and PCL formulations. We implemented a class-reweighted contrastive loss (WCL) in which pairwise similarities between normalized embeddings were optimized via a temperature-scaled log-softmax objective with inverse-frequency sample weighting. This loss aimed to emphasize minority-class samples during pretraining and mitigate the imbalance between hypertensive and normotensive recordings. We benchmarked the standard supervised contrastive (SupCon) \cite{khosla2020supervised}, which optimized the mean log-likelihood of positive pairs within each batch. We then extended this objective by applying class-balanced reweighting (BSCL) based on the effective number of samples to better compensate for imbalanced class distributions. We further explored an extension of the PCL framework by incorporating an angular margin (PCL-AM), inspired by margin-based metric learning methods in face recognition \cite{deng2019arcface,zhang2022incorporating}. This method applied an angular margin to the true-class cosine similarity between embeddings and their corresponding class prototypes before temperature scaling, encouraging larger inter-class angular separation and sharper decision boundaries. Finally, we evaluated the TimeHUT framework \cite{jalali2025learning}, which learned time-series representations by modeling temporal structure through hierarchical uniformity–tolerance balancing of contrastive representations.

\subsection{Model Training}
\label{sec:model-training}
Training and evaluation were performed using stratified five-fold cross-validation. The input scalogram and reciprocal scalogram consisted of 250 time bins and 40 frequency levels. To augment the minority class, we generated up to five independent batches of ten windows for hypertensive subjects, depending on the availability of good-quality windows following data preparation. For the contrastive pretraining approach, the encoder weights were frozen after pre-training to generate embeddings, and subsequently, the final projection head and the classification layer were trained in a standard supervised setting for hypertension detection. All experiments were conducted using Python 3.11.5 with TensorFlow 2.15.0 on a single NVIDIA A100 GPU (40 GB) and 128 GB of host memory.

\section{Results and Analysis} \label{sec:result}

\subsection{Classification Performance}
We evaluated the performance of the proposed hypertension detection framework across different input representations (scalogram, reciprocal scalogram, and a combined multi-view setting) under multiple learning strategies described in section \ref{sec:other_models}. Table \ref{tab:res} summarizes the average performance metrics obtained over all cross-validation folds. To facilitate clinically interpretable evaluation and enable fair comparison, the results of all experiments were reported using an adjusted decision threshold set to achieve an operating point corresponding to a sensitivity of 0.80 \cite{unal2017defining,nahm2022receiver}. 

When using the scalogram representation, the baseline HAN reached an AUROC of 0.71 and an F1 score of 0.34, with a specificity of 0.43 and a balanced accuracy of 0.61 at the selected operating point. Adding CL to the pipeline slightly improved the overall balance of the results by increasing the F1 score to 0.37 and the balanced accuracy to 0.64 while maintaining a comparable AUROC. Incorporating the PCL learning objective produced similar behavior, raising the AUROC to 0.72 and the F1 score to 0.36. Interestingly, having the reciprocal scalogram as the sole input increased the overall discriminative ability. The baseline HAN in this setting achieved an AUROC of 0.78 and a F1 score of 0.41, a lift of roughly seven percentage points over the forward scalogram, while maintaining the fixed sensitivity level. Similarly, the inclusion of CL or PCL frameworks yielded improved and more balanced performance. The PCL-enhanced model achieved the highest F1 score (0.44), specificity (0.60), balanced accuracy (0.70), and accuracy (0.64) within this group, whereas the CL configuration attained the highest AUROC (0.79).

\begin{figure}[ht]
    \centering
    \includegraphics[width=\linewidth]{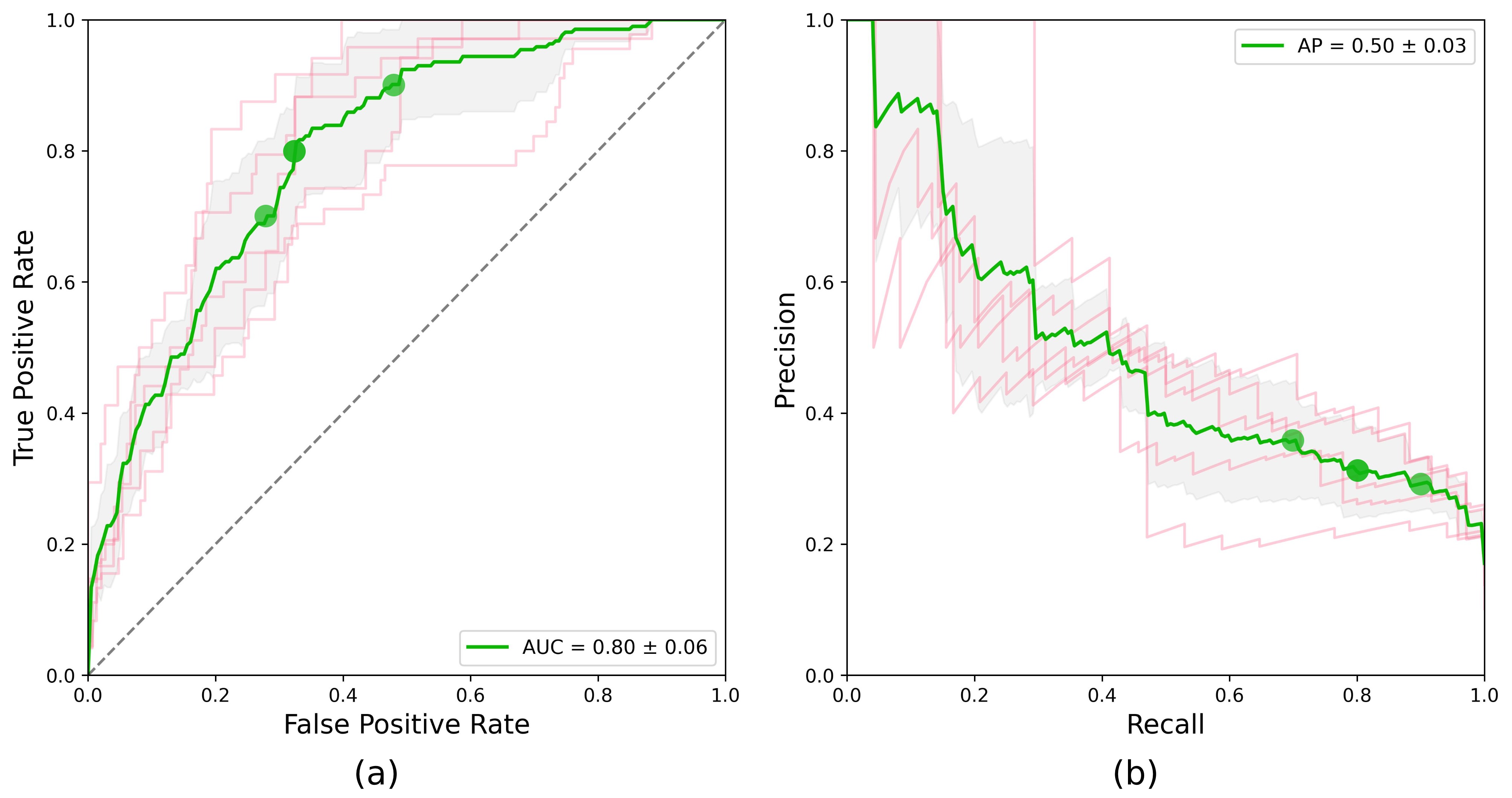}\\
    \caption{Model performance across cross-validation folds. (a) Receiver operating characteristic (ROC) curves for each fold (pink) and their mean curve (green). (b) Precision–recall (PR) curves for each fold (pink) and their mean curve (green). The shaded areas represent the standard deviation among folds, and dots mark the operating points corresponding to sensitivities of 0.70, 0.80, and 0.90.}
    \label{fig:roc}   
\end{figure}

In the multi-view setting, which jointly integrates both forward and reciprocal scalograms, the HAN model achieved an AUROC of 0.78, an F1 score of 0.41, and a specificity of 0.57 at a fixed sensitivity of 0.80. Additional learning strategies produced small but consistent improvements. Among all configurations, the proposed AutoHyPE approach delivered the best and most stable overall performance, with an AUROC of 0.80, an F1 of 0.45, and a balanced accuracy of 0.72. Notably, AutoHyPE increased specificity (0.64), NPV (0.94), and balanced accuracy (0.72) while operating at the same sensitivity level, resulting in a more balanced and reliable decision boundary across classes. Figure \ref{fig:roc} visualizes the receiver operating characteristic (ROC) and precision–recall (PR) curves for this approach. For completeness and comparison, we additionally reported the performance of the enhanced HAN model with PCL for different input representations and reference models at alternative operating points corresponding to sensitivities of 0.70 and 0.90 in Tables \ref{tab:opp}.

\begin{table*}[ht]
\centering
\begin{minipage}{\textwidth}
    \centering
    \fontsize{9pt}{11pt}\selectfont
    \setlength\tabcolsep{4pt}
    \caption{Performance comparison of the proposed model under different development strategies and input representations (scalogram, reciprocal scalogram, and multi-view). Table reports mean $\pm$ standard deviation, and underlined values indicate statistically significant improvement (paired t-test, $p < 0.05$) achieved by AutoHyPE over the corresponding comparison method across cross-validation folds.}
    \label{tab:res}
    \resizebox{\linewidth}{!}{
    \begin{tabular}{lccccccc}
        \toprule
        \bf Model & \bf AUROC & \bf F1 & \bf Sensitivity & \bf Specificity & \bf NPV & \bf BA & \bf Accuracy
        \\
        \midrule
        \rowcolor{gray!10}
        \multicolumn{8}{c}{\textbf{Scalogram}} \\ 
        \midrule
        HAN & $\underline{0.71} \pm 0.06$ & $\underline{0.34} \pm 0.06$ & $0.79 \pm 0.03$ & $\underline{0.43} \pm 0.11$ & $\underline{0.90} \pm 0.05$ & $\underline{0.61} \pm 0.07$ & $\underline{0.49} \pm 0.08$ \\
        ~~+ CL & $\underline{0.71} \pm 0.03$ & $0.37 \pm 0.08$ & $0.80 \pm 0.01$ & $\underline{0.48} \pm 0.11$ & $\underline{0.92} \pm 0.03$ & $0.64 \pm 0.05$ & $0.54 \pm 0.10$ \\
        ~~+ PCL & $\underline{0.72} \pm 0.06$ & $\underline{0.36} \pm 0.04$ & $0.80 \pm 0.01$ & $\underline{0.46} \pm 0.09$ & $\underline{0.91} \pm 0.04$ & $\underline{0.63} \pm 0.05$ & $\underline{0.52} \pm 0.05$ \\
        \midrule
        \rowcolor{gray!10}
        \multicolumn{8}{c}{\textbf{Reciprocal Scalogram}} \\ 
        \midrule
        HAN & $0.78 \pm 0.07$ & $0.41 \pm 0.03$ & $0.80 \pm 0.01$ & $0.58 \pm 0.11$ & $0.92 \pm 0.04$ & $0.69 \pm 0.07$ & $0.62 \pm 0.09$ \\
        ~~+ CL & $0.79 \pm 0.07$ & $0.42 \pm 0.07$ & $0.80 \pm 0.01$ & $0.57 \pm 0.19$ & $0.92 \pm 0.05$ & $0.68 \pm 0.10$ & $0.61 \pm 0.15$ \\
        ~~+ PCL & $0.77 \pm 0.10$ & $0.44 \pm 0.05$ & $0.80 \pm 0.01$ & $0.60 \pm 0.18$ & $0.92 \pm 0.04$ & $0.70 \pm 0.09$ & $0.64 \pm 0.14$ \\
        \midrule
        \rowcolor{gray!10}
        \multicolumn{8}{c}{\textbf{Multi View}} \\ 
        \midrule
        HAN & $0.78 \pm 0.07$ & $0.41 \pm 0.07$ & $0.80 \pm 0.01$ & $0.57 \pm 0.14$ & $0.92 \pm 0.04$ & $0.68 \pm 0.07$ & $0.61 \pm 0.10$	\\
        ~~+ WCL & $0.77 \pm 0.05$ & $0.41 \pm 0.05$ & $0.80 \pm 0.03$ & $0.58 \pm 0.06$ & $0.93 \pm 0.03$ & $0.69 \pm 0.03$ & $0.62 \pm 0.04$ \\
        ~~+ TimeHUT \cite{jalali2025learning} & $0.77 \pm 0.03$ & $0.42 \pm 0.05$ & $0.80 \pm 0.01$ & $0.58 \pm 0.13$ & $0.93 \pm 0.03$ & $0.69 \pm 0.06$ & $0.62 \pm 0.09$ \\
        ~~+ CL & $0.79 \pm 0.05$ & $0.42 \pm 0.06$ & $0.80 \pm 0.01$ & $0.58 \pm 0.08$ & $0.93 \pm 0.03$ & $0.69 \pm 0.04$ & $0.62 \pm 0.06$	\\
        ~~+ SupCon \cite{khosla2020supervised} & $0.79 \pm 0.04$ & $0.42 \pm 0.05$ & $0.80 \pm 0.01$ & $0.59 \pm 0.07$ & $0.92 \pm 0.04$ & $0.69 \pm 0.05$ & $0.63 \pm 0.05$	\\
        ~~+ BSCL & $0.79 \pm 0.08$ & $0.43 \pm 0.03$ & $0.80 \pm 0.01$ & $0.60 \pm 0.12$ & $0.92 \pm 0.04$ & $0.70 \pm 0.06$ & $0.64 \pm 0.09$	\\
        ~~+ PCL-AM & $0.80 \pm 0.05$ & $0.43 \pm 0.05$ & $0.80 \pm 0.02$ & $0.61 \pm 0.09$ & $0.93 \pm 0.03$ & $0.71 \pm 0.05$ & $0.64 \pm 0.07$ \\
        \midrule
        \rowcolor{green!10}
        ~~+ PCL (AutoHyPE) & $0.80 \pm 0.06$ & $0.45 \pm 0.04$ & $0.80 \pm 0.01$ & $0.64 \pm 0.13$ & $0.94 \pm 0.04$ & $0.72 \pm 0.07$ & $0.67 \pm 0.10$\\
        \bottomrule
    \end{tabular}
    }
    \caption*{{\textit{(Abbreviations, 
    AUROC: Area Under the Receiver Operating Characteristic Curve; NPV: Negative Predictive Value; BA: Balanced Accuracy)}}}
\end{minipage}
\end{table*}

\subsection{Ablation Studies}
Since prior works suggested that temperature scheduling strategies can influence the performance of different CL-based frameworks in certain applications, we assessed several scheduling approaches for the AutoHyPE model, as detailed in Table \ref{tab:t_abl}. In the stepwise and cosine annealing strategies, the temperature oscillates between predefined minimum (0.05) and maximum (0.5) values throughout the training process. In the monotonic increase and monotonic decay schemes, the temperature is smoothly ramped between these bounds in a single direction, and in the adaptive setting the temperature is treated as a learnable parameter. Overall, all schedulers yielded broadly comparable performance, but the fluctuating schedules (stepwise and cosine annealing) tended to underperform relative to the baseline configuration. The adaptive and monotonic decay schedules produced the strongest results among the strategies, but their performance remained slightly below that achieved using the fixed temperature value employed in our main experiments. Furthermore, Table \ref{tab:abl} summarizes the results of the architectural ablation study and state-of-the-art reference models. Removing the window-level encoder caused the largest degradation of discriminative power, reducing the AUROC to 0.71. Eliminating the convolutional feature extractor led to a pronounced decline in the model’s ability and reliability in detecting normotensive cases, with specificity falling to 0.55 and the NPV to 0.75. The architecture was least affected when the hierarchical structure was collapsed or when the sentence encoder was removed, although even in these cases the AUROC and F1 scores were approximately two and four percentage points lower, respectively, compared with the original model. The evaluated reference models achieved nearly comparable performance, but below that of the baseline HAN model; InceptionTime reached an AUROC of 0.74 with an F1 score of 0.40, while ROCKET achieved an AUROC of 0.73 and an F1 score of 0.39.

To examine whether additional complementary views could further enhance performance, we incorporated a third input view derived from the instantaneous phase of the complex-valued wavelet coefficients obtained from the CWT. This phase representation captured the relative timing and alignment of oscillatory components across scales and time, providing information distinct from amplitude-based features. The results showed that adding the phase view improved the performance of the baseline approach; however, it did not lead to additional gains for the AutoHyPE variant (Table (\ref{tab:phase})). Moreover, to analyze the effect of input modality, we replaced the scalogram representations with spectrograms while keeping all other settings unchanged. The results of this ablation, reported in Table \ref{tab:spect}, showed considerably lower performance compared to the scalogram across different approaches.

\begin{table*}[ht]
\centering
\begin{minipage}{\textwidth}
    \centering
    \fontsize{9pt}{11pt}\selectfont
    \setlength\tabcolsep{4pt}
    \caption{Ablation study of temperature parameter ($\tau$) scheduling strategies for the AutoHyPE model. Reported values represent mean $\pm$ standard deviation across cross-validation folds.}
    \label{tab:t_abl}
    \resizebox{\linewidth}{!}{
    \begin{tabular}{lccccccc}
        \toprule
        \bf Scheduler & \bf AUROC & \bf F1 & \bf Sensitivity & \bf Specificity & \bf NPV & \bf BA & \bf Accuracy
        \\
        \midrule
        Stepwise & $0.76 \pm 0.05$ & $0.40 \pm 0.04$ & $0.80 \pm 0.01$ & $0.54 \pm 0.10$ & $0.92 \pm 0.04$ & $0.67 \pm 0.05$ & $0.59 \pm 0.07$ \\
        Cosine Annealing & $0.77 \pm 0.03$ & $0.41 \pm 0.08$ & $0.80 \pm 0.01$ & $0.57 \pm 0.06$ & $0.93 \pm 0.02$ & $0.69 \pm 0.03$ & $0.61 \pm 0.05$ \\
        Monotonic Increase & $0.79 \pm 0.04$ & $0.43 \pm 0.06$ & $0.80 \pm 0.02$ & $0.61 \pm 0.07$ & $0.93 \pm 0.03$ & $0.71 \pm 0.04$ & $0.65 \pm 0.05$ \\
        Monotonic Decay & $0.80 \pm 0.05$ & $0.44 \pm 0.05$ & $0.80 \pm 0.01$ & $0.61 \pm 0.15$ & $0.92 \pm 0.05$ & $0.71 \pm 0.08$ & $0.65 \pm 0.11$ \\
        Adaptive & $0.80 \pm 0.05$ & $0.44 \pm 0.07$ & $0.80 \pm 0.01$ & $0.63 \pm 0.09$ & $0.93 \pm 0.03$ & $0.71 \pm 0.08$ & $0.66 \pm 0.13$ \\
        \bottomrule
    \end{tabular}
    }
    \caption*{{\textit{(Abbreviations, 
    AUROC: Area Under the Receiver Operating Characteristic Curve; NPV: Negative Predictive Value; BA: Balanced Accuracy)}}}
\end{minipage}
\end{table*}

\subsection{Real-time Deployment Feasibility}
The proposed framework was designed with real-time deployment and integration into the mobile health platform for prenatal care in mind. We conducted an inference-time analysis of the trained AutoHyPE model and additionally converted it to the TensorFlow Lite (TFLite) format without applying quantization, enabling a direct comparison with the original implementation. Following conversion, the TFLite model was evaluated on the same data and achieved results identical to those reported in Table \ref{tab:res}, with no observable performance degradation, while reducing the model size by 84.16\% (from about 2.03 MB to 0.32 MB). The analysis demonstrated a mean total inference time of $47.49 \pm 2.59$ ms for the original model and $28.86 \pm 2.00$ ms for the converted model when generating predictions from ten 3.75-second multi-view input windows on the training hardware configuration described in section \ref{sec:model-training}, corresponding to an approximately 1.65× speedup. Furthermore, using the \textit{safe+natal} mobile health platform \cite{katebi2024edge}, we estimated the model’s edge inference time on a Google Pixel 6a to be $203.60 \pm 13.43$ ms.

\section{Discussion} \label{sec:discussion}
This study provided evidence that fetal 1D-DUS signals contained latent information reflective of maternal hypertensive status, supporting the hypothesis that fetal cardiovascular dynamics respond measurably to maternal–placental hemodynamic stress. Leveraging one of the largest datasets of paired fetal 1D-DUS and maternal blood pressure recordings assembled to date, we showed that the fetal circulation is not merely a passive recipient of maternal physiology, but an active and sensitive reporter of maternal conditions. While 1D-DUS signals were traditionally interpreted through explicit indices or FHRV metrics, our findings suggested that subtler, distributed temporal–spectral patterns encoded clinically meaningful signatures of maternal hypertension.

The developed hierarchical modeling of input waveforms across temporal scales enabled the network to extract intrinsic structures. Short-term windows captured beat-level morphology and rhythmic variability, while longer sequences aggregated these local representations into higher-level patterns that persist over time. Analysis of the architecture demonstrated the effectiveness of this design; specifically, incorporating window-level (short-term), sentence-level (long-term), and hierarchical encodings yielded improvements of 9, 2, and 2 percentage points in AUROC, respectively, for the best-performing training strategy (Table \ref{tab:abl}). Additionally, this design can facilitate knowledge transfer to other tasks, as the trained encoders can be reused as fixed feature extractors or be fine-tuned for a variety of downstream applications.

A central challenge tackled in this study was the long-tailed, highly imbalanced nature of maternal hypertension data, compounded by heterogeneous fetal physiological signals. Hypertensive cases represented a clear minority relative to normotensive visits, a setting in which conventional supervised learning objectives tended to favor the majority class. Within this context, simply increasing model complexity or depth was ineffective and often counterproductive, as it exacerbated overfitting to majority-class patterns while failing to capture underrepresented but clinically meaningful physiological signatures. This imbalance directly affected error profiles; false negatives (missed hypertensive cases) posed a clinical risk by delaying intervention, while false positives could lead to unnecessary referrals, anxiety, and resource strain in already limited healthcare settings, underscoring the importance of achieving balanced model performance. To provide a consistent basis for comparing different approaches and ensure that performance improvements reflect enhanced discrimination rather than shifts in the decision threshold, results were reported at a fixed operating point corresponding to a sensitivity of 0.80, meaning that approximately one out of five hypertensive cases may be missed. In our experiments, the baseline approach exhibited a marked performance asymmetry, with around two times higher sensitivity than specificity and an F1 score of 0.34 (Table \ref{tab:res}). One of our key contributions in response lied in demonstrating the importance of input geometry and the use of multiple views, particularly given network inputs exhibited bimodal boundary distributions characterized by sharp peaks and large flat, low-energy regions. In such cases, dominant high-energy components could overwhelm the learning process, while weaker but informative spectral cues remained underexploited. From an optimization perspective, these regions with low-variance generated weak gradients, leading to saturation in early network layers. By requiring the network to learn features that remained discriminative across both transformations, the RFA strategy also implicitly constrained the hypothesis space complexity and discouraged reliance on spurious, view-specific patterns. This acted as a form of implicit regularization that empirically reduced overfitting. As a result, the multi-view approach improved the F1 score and specificity of the baseline HAN by seven and 14 percentage points, respectively, compared with the single-view scalogram at the same operating point. Incorporating additional views such as phase further improved the baseline performance but did not benefit the contrastive variants, suggesting that phase-related cues were already captured through other views in those settings by the model (Table \ref{tab:phase}). These findings indicated that the multi-view paradigm was broadly applicable to other domains facing similar data imbalance, distribution, and heterogeneity challenges.

The integration of contrastive pretraining, particularly through a prototype-based contrastive objective, further strengthened the learned representations. By organizing embeddings around learnable class-level prototypes, the model achieved more stable and semantically coherent feature spaces, improving the detectability of minority-class patterns while maintaining robustness against false alarms. This was reflected in a four-percentage-point improvement in F1 score for the multi-view approach when pretrained using PCL. Beyond performance gains, these prototypes could serve as proxies for uncertainty estimation in real-world inference, as distances to class centers reflected confidence in model predictions. In contrast, instance-level contrastive methods such as SupCon and CL emphasized discrimination by directly contrasting individual samples within a batch. Although these approaches are often effective in large and balanced datasets, they can be sensitive to batch composition and susceptible to instability in the presence of noisy or weakly aligned positive pairs. Class-reweighted variants, including WCL and BSCL, attempt to mitigate these limitations by amplifying minority-class contributions, yet they still rely on pairwise sample comparisons. In addition, the inclusion of AM and temperature scheduling (Table \ref{tab:t_abl}) did not yield further improvements, suggesting that increased training complexity did not necessarily translate into better performance in this context. By leveraging the complementary strengths of multi-view learning and contrastive pretraining, the proposed AutoHyPE model achieved a more robust decision boundary, as reflected in its superior overall performance.

Importantly, the proposed method is not intended to replace conventional blood pressure measurements, which remain the clinical standard for diagnosing hypertension. Rather, it is positioned as a secondary, supportive screening tool that can be used to cross-check intermittent blood pressure readings and extend toward continuous hypertension risk monitoring using physiological signals, once sufficient validation and appropriate performance criteria are met. Doppler ultrasound devices are inexpensive, portable, and thanks to advanced mobile health platforms, are becoming widely deployed, especially in low-resource environments where community health workers or traditional midwives routinely use them to automatically assess fetal and maternal well-being. This complementary role is particularly notable in such settings where blood pressure is measured infrequently and under suboptimal conditions. Leveraging these existing devices and platforms to extract additional maternal risk signatures requires no changes to hardware or clinical workflow, making our approach a scalable, low-overhead extension for edge-based analysis. Of note, our development pipeline was designed to maximize compatibility with existing platform components, from data preparation and scalogram generation to model training and deployment, facilitating seamless integration alongside current fetal health monitoring models. That said, this approach should still serve as a supportive layer (for example, as a secondary rule-out tool given its high NPV of 0.94 at an operating point corresponding to a sensitivity of 0.80) and continued advocacy for the deployment of blood pressure devices in under-resourced settings remains essential.

Despite the encouraging findings, several limitations should be considered when interpreting this study. External validation was not conducted, as no independent dataset containing fetal 1D-DUS and maternal blood pressure recordings has been available to date. This constraint limited the ability to fully assess the generalizability of the proposed approach across other populations and data acquisition equipment. In addition, the analysis did not account for neonatal complications such as potential FGR, medication use, or other maternal comorbidities, which could influence fetal hemodynamics and perinatal status. Nevertheless, this study revealed a clear relationship between fetal cardiac activity and maternal hypertensive status, which can be captured seamlessly, noninvasively, and at low cost. Building on these results, future works will focus on leveraging the ongoing prospective data collection to further evaluate, analyze, and refine the framework. As additional data becomes available, the approach will be continuously updated and improved through continual learning strategies, enabling it to adapt to evolving data distributions across different populations. A key goal is to embed the matured model directly into the existing fetal-maternal mobile health platform to support real-time, point-of-care decision-making by midwives and frontline healthcare providers. In parallel, we aim to enrich the modeling framework by incorporating additional clinical, multimodal information such as FGR, maternal demographics, medical history, and other risk factors, thereby moving from a single task hypertension detector toward a more comprehensive maternal fetal risk assessment tool. Most importantly, because longitudinal data across multiple gestational visits have already been collected and are available, the framework can be extended from a diagnostic tool to a predictive model capable of estimating the risk of developing hypertensive disorders later in pregnancy. Such an approach could provide substantial clinical value in both clinical and community-based care by supporting earlier identification of high-risk pregnancies and facilitating timely interventions.

\section{Conclusion} \label{sec:conclusion}
In this work, we showed that fetal 1D-DUS signals encoded discernible hemodynamic signatures of maternal hypertension, revealing a meaningful link between fetal cardiovascular dynamics and maternal blood pressure status. The proposed AutoHyPE framework highlighted how representation geometry mattered as much as representation content when learning from complex inputs. By coupling a deep hierarchical attention architecture with a multi-view representation and prototype-based contrastive learning, the framework uncovered distributed, diagnostically relevant patterns in the collected recordings in real-world community settings. This design effectively addressed persistent challenges of extreme class imbalance and biological heterogeneity, for which network depth alone was insufficient to meaningfully extract latent structure. Detailed analysis and ablations demonstrated the critical effect of each main component and the strategies of our method. When deployed at the edge within a prenatal mobile health platform, these contributions could support the premise that fetal physiology may serve as a non-invasive, scalable, and low-cost proxy for continuous maternal hypertension monitoring, providing a supportive risk awareness system that complements standard intermittent blood pressure measurements.

\section*{Acknowledgment}
This work was funded by the Eunice Kennedy Shriver National Institute of Child Health and Human Development (NICHD), grant number 1R21HD084114-01 to GC (Mobile Health Intervention to Improve Perinatal Continuum of Care in Guatemala), NICHD grant number 1R01HD110480 to GC (AI-driven low-cost ultrasound for automated quantification of hypertension, preeclampsia, and IUGR), Google.org AI for the Global Goals Impact Challenge Award to GC, NK and PR. NK is partially funded by a PREHS-SEED award grant K12ESO33593. The content is solely the responsibility of the authors and does not necessarily represent the official views of the National Institutes of Health.

{
\bibliographystyle{unsrt}
\bibliography{egbib}
}

\clearpage
\appendix

\begin{center}
    \maketitle
    \Large
    \textbf{Appendix}
\end{center}

\setcounter{table}{0}
\setcounter{figure}{0}
\setcounter{equation}{0}
\renewcommand{\theequation}{S\arabic{equation}}
\renewcommand{\thetable}{S\arabic{table}}
\renewcommand\thefigure{S\arabic{figure}}

\section{Descriptive Insights into the Collected Dataset}

\begin{table*}[ht]
\centering
\begin{minipage}{\textwidth}
    \centering
    \fontsize{9pt}{11pt}\selectfont
    \setlength\tabcolsep{4pt}
    \caption{Mean and standard deviation of maternal SBP and DBP measurements collected during visits across gestational ages (months 1–9). All values are reported in mmHg.}
    \label{tab:ga_bb}
    \resizebox{0.9\linewidth}{!}{
    \begin{tabular}{lccccccccc}
        \toprule
        \textbf{Gestational Age (Month)} & \textbf{1} & \textbf{2} & \textbf{3} & \textbf{4} & \textbf{5} & \textbf{6} & \textbf{7} & \textbf{8} & \textbf{9} \\
        \midrule 
        SBP (Mean)  & 103.88 & 102.43 & 100.67 & 100.29 & 99.67 & 99.84 & 101.15 & 103.96 & 109.33 \\
        SBP (Std)   & 14.38  & 12.58  & 8.30   & 10.21  & 8.79  & 8.90  & 9.06   & 10.09  & 12.70 \\
        DBP (Mean)  & 64.17  & 63.72  & 62.20  & 60.66  & 60.50 & 61.08 & 62.41  & 65.29  & 69.43 \\
        DBP (Std)   & 6.71   & 7.60   & 6.98   & 7.50   & 6.89  & 6.63  & 7.16   & 7.97   & 9.69  \\
        \bottomrule
    \end{tabular}
    }
    \caption*{{\textit{(Abbreviations, 
    SBP: Systolic Blood Pressure; DBP: Diastolic Blood Pressure)}}}
\end{minipage}
\end{table*}

\begin{figure}[ht]
    \centering
    \includegraphics[width=0.5\linewidth]{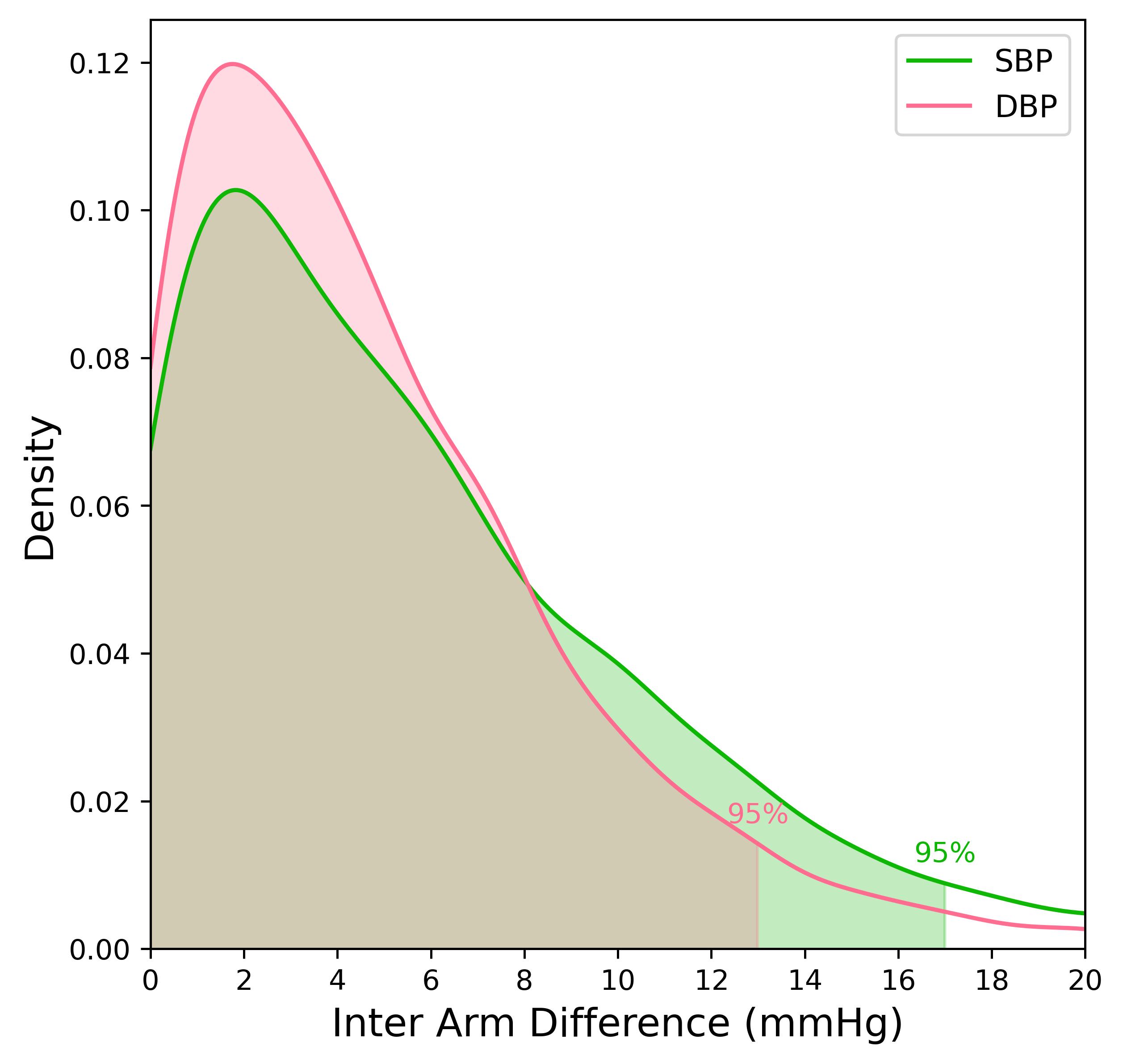}\\
    \caption{Kernel density plots of inter-arm differences for SBP and DBP measurements in the dataset. Both measures display right-skewed distributions, with the majority of differences clustered within lower ranges. The shaded regions highlight the 95th percentile thresholds for each distribution.}
    \label{fig:iad}   
\end{figure}

\section{Multi-View Signal Characterization} \label{sec:char}
To analyze how the information provided by the two views differed in the proposed multi-view representation, we performed a detailed comparative characterization of the scalogram and its reciprocal transformation across hypertensive and normotensive recordings. Although the two views were deterministically related and therefore information-preserving, they induced markedly different statistical and geometric properties in the time–frequency domain, which influenced how discriminative patterns were exposed to a learning model.

\begin{figure}[ht]
    \centering
    \includegraphics[width=0.7\linewidth]{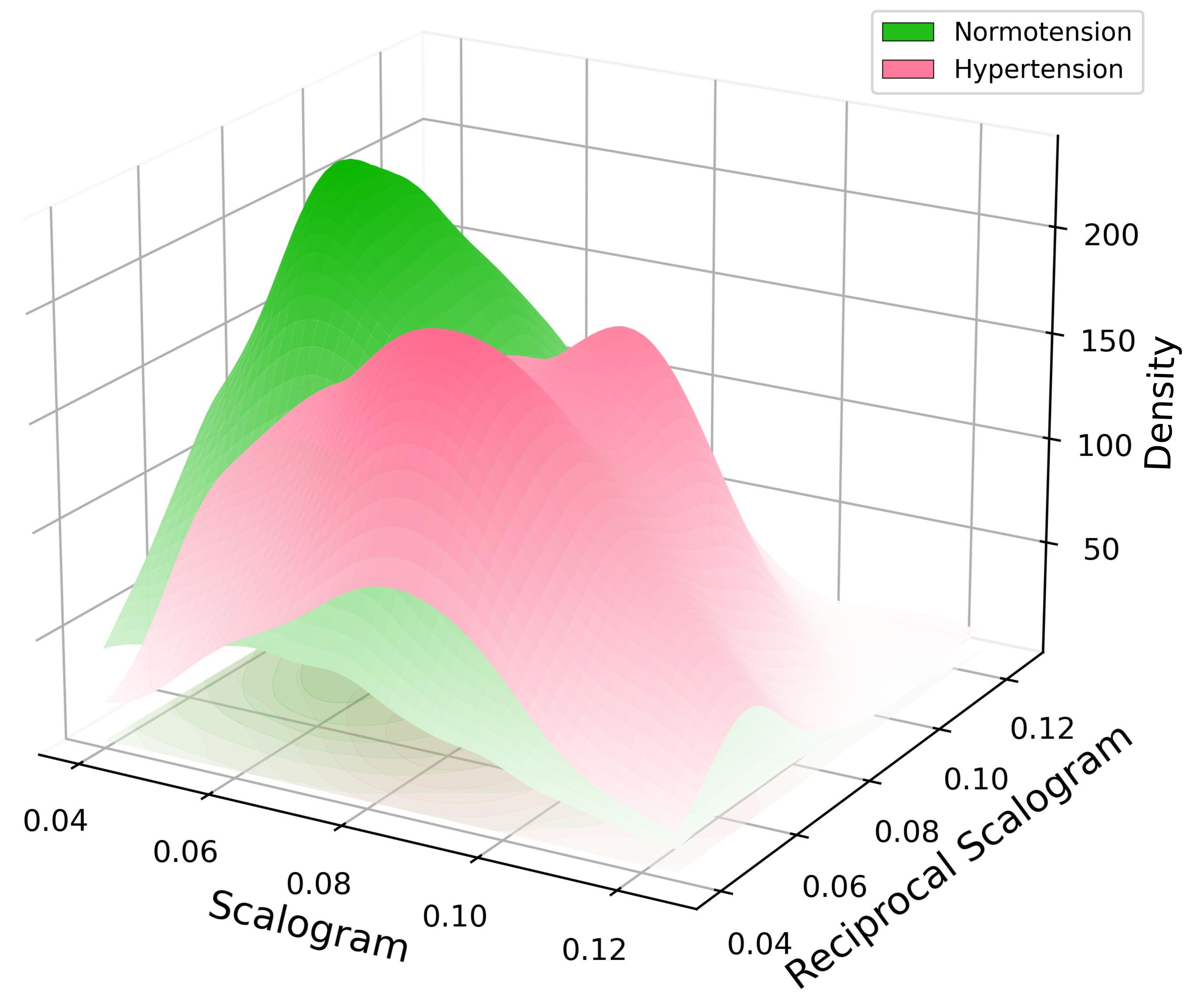}\\
    \caption{3D kernel density plots showing the distribution of mean spectral power values in the scalogram and reciprocal scalogram for each heartbeat chunk within 3.75-second windows for hypertensive and normotensive groups.}
    \label{fig:3d_kdp}   
\end{figure}

We first examined the distribution of spectral energy within 3.75-second windows by computing the mean spectral power across individual heartbeat segments. To achieve this, we employed the AutoFHR model \cite{rafiei2025next,rafiei2025autofhr} to segment the heartbeats within each window. As illustrated in Figure \ref{fig:3d_kdp}, the resulting joint kernel density estimates revealed a nonlinear reparameterization of spectral magnitude. In hypertensive cases, mean heartbeat-level power values tended to be lower in the forward scalogram and higher in the reciprocal representation. In contrast, normotensive recordings exhibited a more concentrated distribution in the scalogram and a more dispersed distribution in the reciprocal view. This redistribution suggested that diagnostically relevant patterns may be emphasized differently across the two representations.

\begin{figure}[ht]
    \centering
    \includegraphics[width=\linewidth]{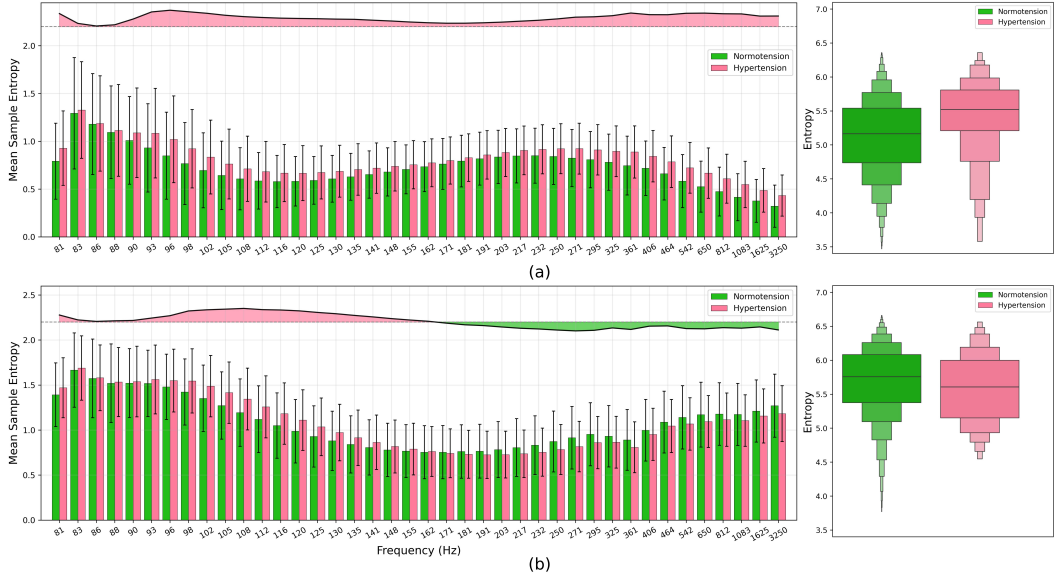}\\
    \caption{Average sample entropy across frequency scales with standard deviation bar highlighted (left) and the distribution of image entropy values (right) for (a) scalograms and (b) reciprocal scalograms of the 3.75-second windows, computed for both hypertensive and normotensive cases. The difference profile between the bar values of the two classes for each scale is overlaid at the top of each panel.}
    \label{fig:entropy}   
\end{figure}

We further quantified structural complexity using entropy-based measures. Sample entropy was computed to assess irregularity and variability of spectral patterns across frequency scales, while image entropy, calculated for each 2D 3.75-second time–frequency window, summarized overall textural complexity. As shown in Figure \ref{fig:entropy}, hypertensive scalograms consistently exhibited higher mean sample entropy across all frequency scales compared to normotensive recordings. In the reciprocal scalogram, however, the trend varied by frequency: hypertensive cases showed higher entropy at lower frequencies but lower entropy at higher frequency bands relative to the normotensive group. Additionally, the standard scalogram demonstrated greater sensitivity to group-level differences in entropy distributions.

\newpage
\section{Additional Results and Ablation Studies}

\begin{table*}[ht]
\centering
\begin{minipage}{\textwidth}
    \centering
    \fontsize{9pt}{11pt}\selectfont
    \setlength\tabcolsep{4pt}
    \caption{Performance comparison of the proposed HAN + PCL approach under different input representations (scalogram, reciprocal scalogram, and multi-view), along with the reference model using the multi-view strategy, at alternative operating points. Values are reported as mean $\pm$ standard deviation.}
    \label{tab:opp}
    \resizebox{\linewidth}{!}{
    \begin{tabular}{lccccccc}
        \toprule
        \bf Approach & \bf AUROC & \bf F1 & \bf Sensitivity & \bf Specificity & \bf NPV & \bf BA & \bf Accuracy
        \\
        \midrule
        \rowcolor{gray!10}
        \multicolumn{8}{c}{\textbf{Operating Point: Sensitivity = 0.70}} \\ 
        \midrule
        Scalogram & $0.71 \pm 0.06$ & $0.37 \pm 0.02$ & $0.70 \pm 0.01$ & $0.58 \pm 0.10$ & $0.90 \pm 0.04$ & $0.64 \pm 0.05$ & $0.67 \pm 0.08$ \\
        ROCKET & $0.73 \pm 0.04$ & $0.40 \pm 0.08$ & $0.70 \pm 0.00$ & $0.66 \pm 0.06$ & $0.92 \pm 0.02$ & $0.68 \pm 0.03$ & $0.67 \pm 0.05$ \\
        InceptionTime & $0.74 \pm 0.07$ & $0.40 \pm 0.10$ & $0.70 \pm 0.00$ & $0.62 \pm 0.12$ & $0.90 \pm 0.04$ & $0.66 \pm 0.06$ & $0.63 \pm 0.10$ \\
        Reciprocal Scalogram & $0.78 \pm 0.07$ & $0.47 \pm 0.10$ & $0.70 \pm 0.01$ & $0.69 \pm 0.18$ & $0.90 \pm 0.05$ & $0.69 \pm 0.09$ & $0.70 \pm 0.10$ \\
        Multi View & $0.80 \pm 0.06$ & $0.47 \pm 0.02$ & $0.70 \pm 0.01$ & $0.74 \pm 0.08$ & $0.92 \pm 0.03$ & $0.72 \pm 0.04$ & $0.73 \pm 0.07$ \\
        \midrule
        \rowcolor{gray!10}
        \multicolumn{8}{c}{\textbf{Operating Point: Sensitivity = 0.90}} \\ 
        \midrule
        Scalogram & $0.71 \pm 0.06$ & $0.35 \pm 0.06$ & $0.91 \pm 0.01$ & $0.33 \pm 0.09$ & $0.94 \pm 0.03$ & $0.62 \pm 0.05$ & $0.43 \pm 0.06$ \\
        ROCKET & $0.73 \pm 0.04$ & $0.34 \pm 0.07$ & $0.90 \pm 0.01$ & $0.32 \pm 0.11$ & $0.94 \pm 0.02$ & $0.61 \pm 0.05$ & $0.42 \pm 0.09$ \\
        InceptionTime & $0.74 \pm 0.07$ & $0.36 \pm 0.11$ & $0.91 \pm 0.01$ & $0.32 \pm 0.25$ & $0.89 \pm 0.10$ & $0.61 \pm 0.12$ & $0.43 \pm 0.20$ \\
        Reciprocal Scalogram & $0.78 \pm 0.07$ & $0.41 \pm 0.05$ & $0.91 \pm 0.01$ & $0.48 \pm 0.16$ & $0.96 \pm 0.02$ & $0.69 \pm 0.07$ & $0.56 \pm 0.12$ \\
        Multi View & $0.80 \pm 0.06$ & $0.44 \pm 0.02$ & $0.90 \pm 0.01$ & $0.54 \pm 0.15$ & $0.96 \pm 0.02$ & $0.72 \pm 0.07$ & $0.61 \pm 0.11$ \\
        \bottomrule
    \end{tabular}
    }
    \caption*{{\textit{(Abbreviations, 
    AUROC: Area Under the Receiver Operating Characteristic Curve; NPV: Negative Predictive Value; BA: Balanced Accuracy)}}}
\end{minipage}
\end{table*}

\begin{table*}[ht]
\centering
\begin{minipage}{\textwidth}
    \centering
    \fontsize{9pt}{11pt}\selectfont
    \setlength\tabcolsep{4pt}
    \caption{Comparison of architectural ablations and reference models’ performance. Architectural ablation study evaluates the contribution of different components and configurations within the HAN model for the best performing training strategy. Each row reports the performance of the model after removing/modifying a specific module, and the values represent mean $\pm$ standard deviation across cross-validation folds.}
    \label{tab:abl}
    \resizebox{\linewidth}{!}{
    \begin{tabular}{lccccccc}
        \toprule
        \bf Ablation (reference model) & \bf AUROC & \bf F1 & \bf Sensitivity & \bf Specificity & \bf NPV & \bf BA & \bf Accuracy
        \\
        \midrule
        w/o Window-level Encoder & $0.71 \pm 0.05$ & $0.36 \pm 0.10$ & $0.79 \pm 0.01$ & $0.45 \pm 0.22$ & $0.89 \pm 0.07$ & $0.62 \pm 0.11$ & $0.50 \pm 0.19$ \\
        (ROCKET \cite{dempster_etal_2020}) & $0.73 \pm 0.04$ & $0.39 \pm 0.10$ & $0.79 \pm 0.01$ & $0.56 \pm 0.07$ & $0.93 \pm 0.02$ & $0.68 \pm 0.04$ & $0.59 \pm 0.07$ \\
        w/o Recurrent Modules & $0.74 \pm 0.07$ & $0.39 \pm 0.04$ & $0.78 \pm 0.04$ & $0.58 \pm 0.10$ & $0.92 \pm 0.04$ & $0.68 \pm 0.06$ & $0.61 \pm 0.08$ \\
        (InceptionTime \cite{IsmailFawaz2020inceptionTime}) & $0.74 \pm 0.07$ & $0.40 \pm 0.10$ & $0.80 \pm 0.01$ & $0.52 \pm 0.18$ & $0.91 \pm 0.07$ & $0.66 \pm 0.09$ & $0.57 \pm 0.14$ \\
        w/o Window-level Attention & $0.75 \pm 0.09$ & $0.40 \pm 0.06$ & $0.78 \pm 0.03$ & $0.58 \pm 0.13$ & $0.92 \pm 0.04$ & $0.68 \pm 0.07$ & $0.62 \pm 0.11$ \\
        w/o Hierarchical Attentions & $0.77 \pm 0.05$ & $0.40 \pm 0.08$ & $0.78 \pm 0.04$ & $0.58 \pm 0.12$ & $0.92 \pm 0.04$ & $0.68 \pm 0.06$ & $0.62 \pm 0.09$	\\
        w/o Sequence-level Encoder & $0.78 \pm 0.06$ & $0.40 \pm 0.05$ & $0.77 \pm 0.05$ & $0.61 \pm 0.04$ & $0.92 \pm 0.04$ & $0.69 \pm 0.05$ & $0.64 \pm 0.04$ \\
        w/o Convolutional Feature Extractor & $0.78 \pm 0.04$ & $0.41 \pm 0.06$ & $0.79 \pm 0.13$ & $0.55 \pm 0.29$ & $0.75 \pm 0.38$ & $0.67 \pm 0.09$ & $0.60 \pm 0.20$ \\
        Collapsed Hierarchy & $0.78 \pm 0.07$ & $0.42 \pm 0.04$ & $0.80 \pm 0.01$ & $0.60 \pm 0.12$ & $0.92 \pm 0.04$ & $0.69 \pm 0.06$ & $0.64 \pm 0.09$ \\
        \bottomrule
    \end{tabular}
    }
    \caption*{{\textit{(
    Abbreviations, 
    AUROC: Area Under the Receiver Operating Characteristic Curve; NPV: Negative Predictive Value; BA: Balanced Accuracy)}}}
\end{minipage}
\end{table*}


\begin{table*}[ht]
\centering
\begin{minipage}{\textwidth}
    \centering
    \fontsize{9pt}{11pt}\selectfont
    \setlength\tabcolsep{4pt}
    \caption{Performance comparison of the proposed model under different developmental strategies using three-view input representation (scalogram + reciprocal scalogram + phase). Results show mean $\pm$ standard deviation across cross-validation folds.}
    \label{tab:phase}
    \resizebox{\linewidth}{!}{
    \begin{tabular}{lccccccc}
        \toprule
        \bf Model & \bf AUROC & \bf F1 & \bf Sensitivity & \bf Specificity & \bf NPV & \bf BA & \bf Accuracy
        \\
        \midrule 
        HAN & $0.79 \pm 0.06$ & $0.43 \pm 0.08$ & $0.80 \pm 0.02$ & $0.60 \pm 0.12$ & $0.92 \pm 0.05$ & $0.70 \pm 0.07$ & $0.63 \pm 0.10$ \\
       ~~+ CL & $0.79 \pm 0.05$ & $0.41 \pm 0.05$ & $0.80 \pm 0.01$ & $0.58 \pm 0.11$ & $0.93 \pm 0.03$ & $0.69 \pm 0.06$ & $0.62 \pm 0.08$ \\
        ~~+ PCL (AutoHyPE) & $0.79 \pm 0.07$ & $0.44 \pm 0.06$ & $0.80 \pm 0.01$ & $0.62 \pm 0.11$ & $0.93 \pm 0.04$ & $0.71 \pm 0.05$ & $0.65 \pm 0.08$ \\
        \bottomrule
    \end{tabular}
    }
    \caption*{{\textit{(Abbreviations, 
    AUROC: Area Under the Receiver Operating Characteristic Curve; NPV: Negative Predictive Value; BA: Balanced Accuracy)}}}
\end{minipage}
\end{table*}



\begin{table*}[ht]
\centering
\begin{minipage}{\textwidth}
    \centering
    \fontsize{9pt}{11pt}\selectfont
    \setlength\tabcolsep{4pt}
    \caption{Performance of the proposed model under different developmental strategies using the multi-view spectrogram input representation. Results show mean $\pm$ standard deviation across cross-validation folds.}
    \label{tab:spect}
    \resizebox{\linewidth}{!}{
    \begin{tabular}{lccccccc}
        \toprule
        \bf Model & \bf AUROC & \bf F1 & \bf Sensitivity & \bf Specificity & \bf NPV & \bf BA & \bf Accuracy
        \\
        \midrule
        HAN & $0.65 \pm 0.05$ & $0.34 \pm 0.10$ & $0.81 \pm 0.10$ & $0.39 \pm 0.23$ & $0.73 \pm 0.37$ & $0.60 \pm 0.07$ & $0.46 \pm 0.17$ \\
        ~~+ CL & $0.72 \pm 0.03$ & $0.37 \pm 0.07$ & $0.79 \pm 0.01$ & $0.51 \pm 0.04$ & $0.92 \pm 0.03$ & $0.65 \pm 0.02$ & $0.56 \pm 0.03$ \\
        ~~+ PCL (AutoHyPE) & $0.72 \pm 0.03$ & $0.36 \pm 0.08$ & $0.79 \pm 0.01$ & $0.50 \pm 0.07$ & $0.92 \pm 0.03$ & $0.64 \pm 0.03$ & $0.54 \pm 0.06$ \\

        \bottomrule
    \end{tabular}
    }
    \caption*{{\textit{(Abbreviations, 
    AUROC: Area Under the Receiver Operating Characteristic Curve; NPV: Negative Predictive Value; BA: Balanced Accuracy)}}}
\end{minipage}
\end{table*}



\clearpage

\end{document}